\documentclass[aps,prb,amsmath,amssymb,twocolumn,superscriptaddress,showpacs,longbibliography]{revtex4-2}

\usepackage{graphicx}
\usepackage{dcolumn}
\usepackage{bm}
\usepackage{layout}
\usepackage{lipsum}
\usepackage{amsmath}

\renewcommand{\vec}[1]{\bf{#1}}

\usepackage[colorlinks=true, citecolor=blue, linkcolor=blue, urlcolor=blue]{hyperref}

\begin{document}

\preprint{APS/123-QED}

\title{Floquet engineering nearly flat bands through\\quantum-geometric light-matter coupling with surface polaritons}

\newcommand{\affiliationRWTH}{
Institut f\"ur Theorie der Statistischen Physik, RWTH Aachen University and JARA-Fundamentals of Future Information Technology, 52056 Aachen, Germany
}
\newcommand{\affiliationMPSD}{
Max Planck Institute for the Structure and Dynamics of Matter, Center for Free-Electron Laser Science (CFEL), Luruper Chaussee 149, 22761 Hamburg, Germany
}
\newcommand{\affiliationBremen}{
Institute for Theoretical Physics and Bremen Center for Computational Materials Science, University of Bremen, 28359 Bremen, Germany
}

\author{Mikołaj Walicki}
\affiliation{%
 Institute of Theoretical Physics, Faculty of Physics, University of Warsaw, Pasteura 5, PL-02093 Warsaw, Poland
}%
\author{Christian J. Eckhardt}
\affiliation{\affiliationMPSD}
\affiliation{\affiliationRWTH}

\author{Michael A. Sentef}
\affiliation{\affiliationBremen}
\affiliation{\affiliationMPSD}
\email{sentef@uni-bremen.de}

\date{\today}

\begin{abstract}
We investigate Floquet engineering in a sawtooth chain -- a minimal model hosting a nearly flat band endowed with nontrivial quantum geometry -- coupled to driven surface polaritons. In this paradigmatic flat band model, light-matter coupling to a flat band is enabled by quantum geometry despite the vanishing band velocity and band curvature. We show that light polarization and finite momentum transfer in polaritonic settings provide sufficient tunability to flatten or unflatten bands, with sometimes drastic band structure modifications beyond what is attainable with laser pulses in free space. Possible implications for light-driven phenomena in prototypical flat-band moir\'e or kagome materials are discussed.
\end{abstract}

\maketitle
\section{Introduction}
The design of quantum materials through light-matter coupling is a blossoming research field \cite{basov_towards_2017,de_la_torre_colloquium_2021} offering exciting opportunities for creating new states of matter with prospects for applications in optoelectronic devices. Specifically, progress in tailor-made time-resolved dynamics with strong and short laser pulses offers opportunities in the realms of photo-induced phase transitions as well as dressed states in quasi-time-periodic fields, known as Floquet states \cite{rudner_band_2020}. On the other hand, the use of quantum fluctuations of light and their strong coupling to matter excitations \cite{frisk_kockum_ultrastrong_2019,garcia-vidal_manipulating_2021} in cavity quantum materials is an adjacent field of research with increasing activity \cite{schlawin_cavity_2022,bloch_strongly_2022}, as highlighted by groundbreaking experimental demonstrations of cavity-modified quantum Hall effects \cite{appugliese_breakdown_2022} and charge-density wave transitions \cite{jarc_cavity-mediated_2023}.

In Floquet materials science, one of the major hurdles towards achieving breakthrough results across larger classes of materials lies in the need for strong fields coherently dressing the matter states before heating and dissipation start to dominate the physics \cite{aeschlimann_survival_2021}. While this has recently been overcome in several instances, with successful demonstrations of Floquet engineering across different quantum materials platforms \cite{wang_observation_2013,mahmood_selective_2016,mciver_light-induced_2020,shan_giant_2021,zhou_pseudospin-selective_2023,ito_build-up_2023,zhang_light-induced_2024,merboldt_observation_2024,choi_direct_2024,bielinski_revealing_2024}, it remains a challenge to drive materials to the required field strengths for Floquet engineering effects to become sizeable. Hence it is desirable to propose alternative schemes that require lower laser intensities by employing enhanced light-matter coupling. A second limitation of Floquet engineering in free space is that optical or mid-infrared fields usually have wavelengths well above the atomic scale, therefore restricting the scope of light-matter dressing to long-wavelength, small-wavevector regimes in which the dipole approximation typically holds. By contrast, the emerging variety of cavity-like quantum-electrodynamical surroundings with tunable polariton mode dispersions \cite{basov_polariton_2021,herzig_sheinfux_high-quality_2024,kipp_cavity_2024} opens new opportunities to overcome the apparent discrepancy of length and momentum scales between light and matter. Below we specifically consider a cavity setting with surface polaritons, instead of more conventional Fabry-Perot type cavities, precisely because this allows us to make use of larger effective couplings, due to compressed effective mode volumes, and finite momentum transfer between light and matter. We note that cavity quality factors are of less importance in this context compared to other situations, e.g., in atomic and molecular physics.

We target one specific goal: We wish to dress electronic bands in a minimal model for flat band physics. Flat bands with non-trivial quantum geometry can lead to interesting strongly correlated electron phenomena, as evidenced by the blossoming fields of moir\'e heterostructures \cite{cao_unconventional_2018,kennes_moire_2021} and kagome materials \cite{neupert_charge_2022}. General wisdom suggests that Floquet engineering such bands is difficult since nearly flat bands have almost vanishing band velocity and curvature, thus rendering classical paramagnetic and diamagnetic light-matter couplings small. However, their non-trivial quantum geometry can endow nearly flat bands with quantum-geometric light-matter couplings, which in turn provide a handle for Floquet engineering \cite{topp_light-matter_2021}.

We show through model calculations that flexible control over the electronic structure in a sawtooth chain is enabled through its coupling to laser-driven surface polaritons. The degree of flattening of a nearly flat band can be controlled via the choice of light polarization and the range of wave vectors involved in the dressing scheme. This flexible light-matter control scheme can be viewed as a stepping stone for achieving control over the plethora of emergent phases that arise in flat-band materials.

\section{Model and methods}
\begin{figure*}
    \centering
    \includegraphics{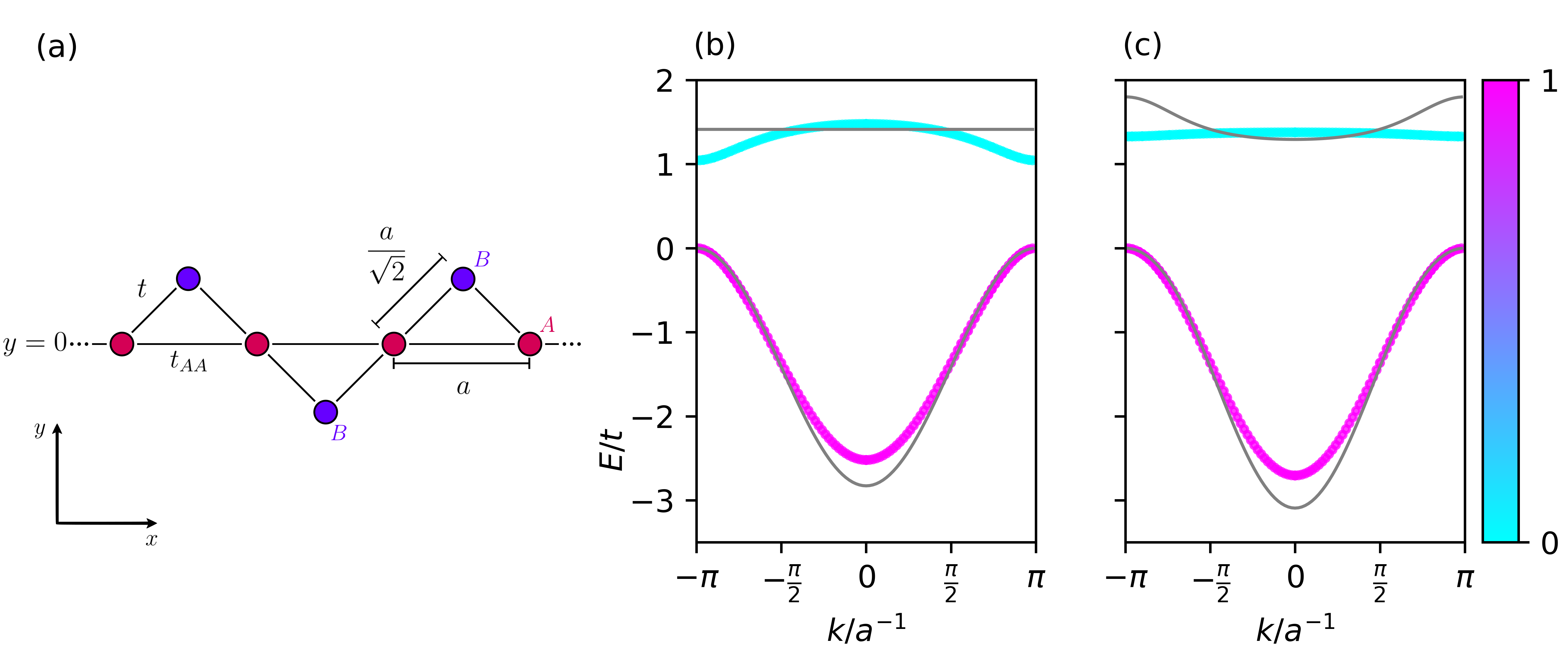}
    \caption{(a) Sawtooth chain with hopping matrix elements $t$ between A and B sites, and $t_{AA}$ for hopping on the A sublattice. 
    (b, c) Band unflattening (b) and flattening (c) for homogeneous ($q=0$) field polarized along $y$ (``perpendicular field'') with amplitude $A_0 = 0.75$. Here $t_{AA} = \frac{t}{\sqrt2}$ in (b) and $t_{AA} = 0.9 t$ in (c). The grey lines show the energy spectrum in the absence of the electromagnetic field. The band coloring indicates the projection of the driven bands onto the undriven eigenstates, with 1 denoting 100\% share of lower bare band and 0 denoting 100\% share of the upper bare band.}
    \label{fig:flat}
\end{figure*}
We investigate the properties of flat bands with non-trivial geometry in a sawtooth chain with lattice constant $a$ (Fig.~\ref{fig:flat}(a)) and Hamiltonian
\begin{equation}
\begin{aligned}
H=- \sum_{i} (t c_{i,A}^\dagger c_{i,B} + tc_{i+1,A}^\dagger c_{i,B} + t_{AA} c_{i+1,A}^\dagger c_{i,A}) + H.c.
\end{aligned}
\end{equation}
Here $t$ denotes inter-sublattice hopping, while $t_{AA}$ denotes intra-sublattice hopping on the A sublattice, as depicted in Fig.~\ref{fig:flat}(a); the canonical fermionic operators $c_{i,a}^\dagger$ create an electron on site $i$ and sublattice $a \in \{A,B\}$, with corresponding annihilation operator $c_{i,a}$. Transformation to momentum space $k$ yields
\begin{equation}
    \begin{aligned}
        H&=- \sum_{k} (h_{AB}(k)c_{k,A}^\dagger c_{k,B} + h_{AA}(k)c_{k,A}^\dagger c_{k,A}) + H.c.,
    \end{aligned}
\end{equation}
where $h_{ab}(k)$ are matrix elements of the $2x2$ Hamiltonian blocks
\begin{equation}
    h(k) = -\begin{pmatrix}
        2t_{AA}\cos ka & t(1+e^{-ika})\\
        t(1+ e^{ika}) & 0\\
    \end{pmatrix}.
\end{equation}
The blocks are then diagonalized to obtain a band structure with eigenvalues
\begin{equation}
    \begin{aligned}
        \varepsilon_1(k)&=-t_{AA}\cos ka -\sqrt{t_{AA}^2\cos ka+ 2t^2(1+\cos ka)},\\
        \varepsilon_2(k)&=-t_{AA}\cos ka +\sqrt{t_{AA}^2\cos ka+ 2t^2(1+\cos ka)}.
    \end{aligned}
\end{equation}
By tuning the ratio $t_{AA}/t$ one can tune the degree of flatness of the upper band. For $t_{AA} = \frac{t}{\sqrt{2}}$ the upper band becomes exactly flat:
\begin{equation}
    \begin{aligned}
        \varepsilon_1(k)&=-\sqrt{2}t(\cos{ka}+1),\\
        \varepsilon_2(k)&=\sqrt{2}t.
    \end{aligned}
\end{equation}
While the model is quasi-one-dimensional, the real-space sawtooth geometry of the chain is crucial for its interaction with light. Here we choose the vector between sites A and B to be ${\bf r_b - r_a} =\frac12 (1,\:1,\:0)^T a$, as depicted in Fig.~\ref{fig:flat}(a).

The system is coupled to a light field given by a vector potential ${\bf A}({\bf r})$ (suppressing its possible time dependence for brevity) via the standard Peierls substitution
\begin{equation}
\begin{aligned}
    H &= \sum_{ab\in \{A,B\}}\sum_{ij} \left(-t_{ab}(i,j)\right)c_{i,a}^\dagger c_{j,b}e^{i\int_{R_j,b}^{R_{i,a}}dr'_\mu A^\mu({\bf r'})}.
\end{aligned}
\end{equation}
This can be approximated by assuming that the wavelength of light is much longer than the lattice constant of the chain, simplifying the integral in the exponent. The vector potential is then expanded as a sum of modes containing different momentum ${\bf q}$ vectors, yielding a perturbative expression for light-matter coupling (here presented up to second order),
\begin{equation}
\begin{aligned}
H&=\sum_{ab\in \{A,B\}}\sum_{\bf k}\{c_{{\bf k},a}^\dagger c_{{\bf k},b}h_{ab}({\bf k}) \\&+ \sum_{\bf q}c_{{\bf k},a}^\dagger c_{{\bf k+q},b}A_{\bf q}^\mu \partial_{k_\mu}h_{ab}({\bf k+\frac{q}{2}})
\\&+ \frac12\sum_{\bf qq'}[c_{{\bf k},a}^\dagger c_{{\bf k+q+q'},b}
 A_{\bf q}^\mu A_{\bf q'}^{\nu}\partial_{k_\mu}\partial_{k_\nu}h_{ab}({\bf k+\frac{q}{2} + \frac{{\bf q'}}{2}})
\\&+ c_{{\bf k},a}^\dagger c_{{\bf k+q-q'},b}
 A_{\bf q}^\mu A_{\bf q'}^{\nu *}\partial_{k_\mu}\partial_{k_\nu}h_{ab}({\bf k+\frac{q}{2} - \frac{ q'}{2}})] \} +H.c.   
\end{aligned}
\label{eq:curvature}
\end{equation}
This expression makes it explicit that flat bands can couple to light despite their vanishing band velocity, since the momentum derivatives act on the Hamiltonian matrix elements, not just on the energy eigenvalues. The potential light-matter coupling in a flat band is thus a result of non-trivial quantum geometry, as already discussed in Ref.~\onlinecite{topp_light-matter_2021}.

In the following we wish to drive the light modes that have been kept rather general up to this point. We consider a time-dependent vector potential of the form
\begin{equation}
{\bf A}({\bf r},t) = \sum_{\bf q} {\bf A}_q\cos({\bf qr}-\omega t),
\label{eq:tdepvector}
\end{equation}
and assume that the frequency of the drive is high compared to bandwidth of the nearly flat band, but small compared to the gap between the flat band and the dispersive band. It is anticipated that such a configuration is realistically achievable specifically for fields stemming from surface polaritons, which are typically in the few terahertz range, well below electronic hopping parameters for most materials. In this limit the time dependence can be averaged by time averaging the Peierls phase factor,
\begin{equation}
\begin{aligned}
    &\langle e^{i\int_{\bf r_i}^{\bf r_j}d{\bf r}{\bf A}({\bf r},t)}\rangle _t \\&\approx
    \frac{1}{T}\int_0^{T}dt e^{i\sum_q{\bf A}_q\cos({\frac{\bf (r_j+r_i)}{2}q}-\omega t)({\bf r_j-r_i})}\\
    &=J_0({\bf A}(\frac{{\bf r_j+r_i}}{2})({\bf r_j-r_i}))\\
    &=1 - \frac14[{\bf A}(\frac{{\bf r_j+r_i}}{2})({\bf r_j-r_i})]^2 + O({\bf A} ^4) ,
    \label{eq:peierls}
\end{aligned}
\end{equation}
where $J_0$ is the zeroth order Bessel function, which has been Taylor-expanded in the last step. One can arrive at the same result by averaging the amplitudes $A^\mu_q$ in (\ref{eq:curvature}). 

Typically, field amplitudes required to induce significant changes in the band structure are rather large, as can be seen by inspecting the renormalized Peierls phase factor in Eq.~\ref{eq:peierls}. The second order term in the last line of the equation suggests that vector potentials of the size of a significant fraction of the Brillouin zone are needed, since the term ${\bf r_j-r_i}$ is on the order of the lattice constant, whose inverse governs the Brillouin zone size. Such enormous field strengths are not attainable in the far field of an optical or terahertz laser drive. To get around this constraint we therefore focus on surface modes, since they allow for the required field enhancement. 

\begin{figure*}[htp!]
    \centering
    \includegraphics{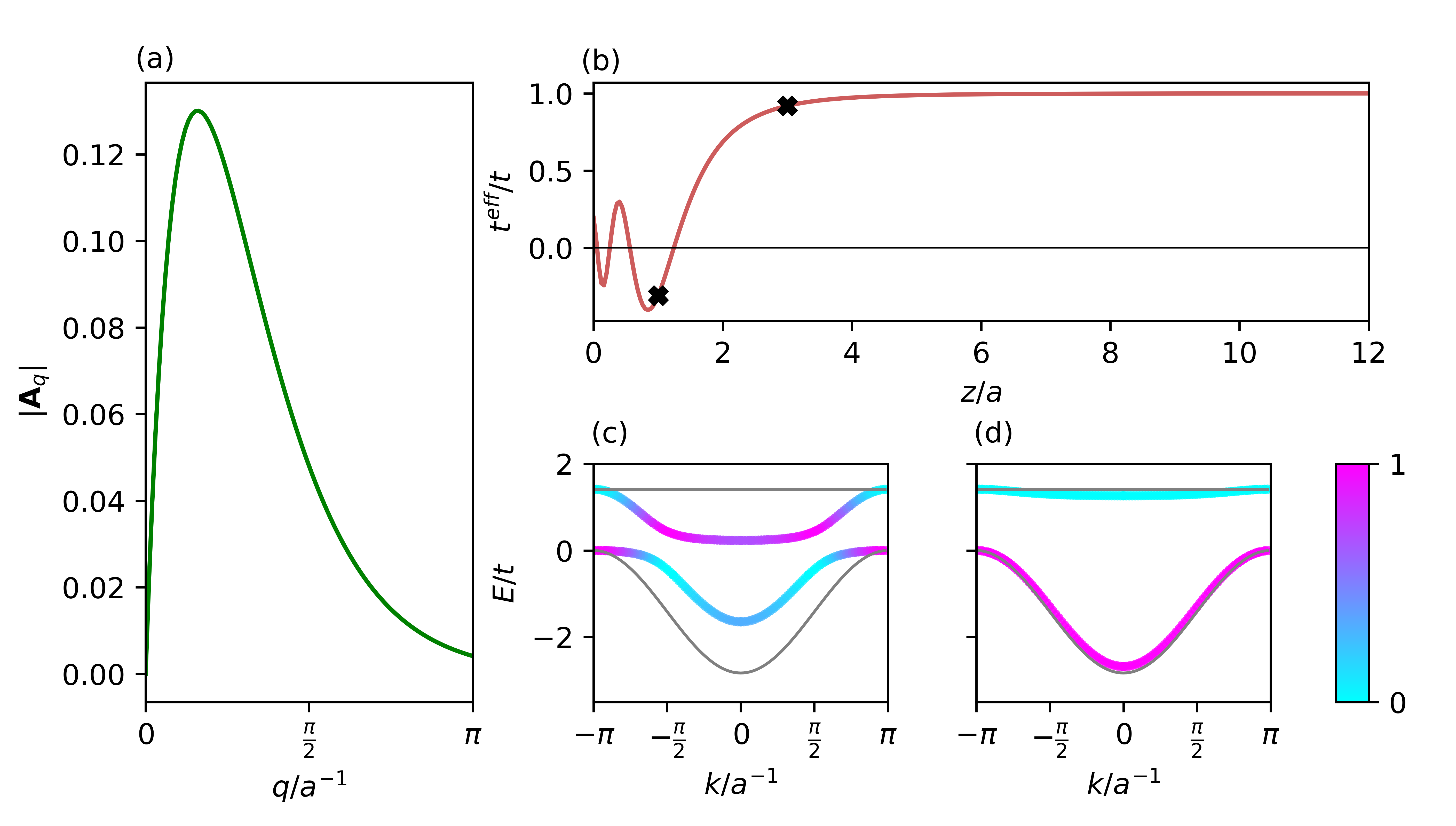}
    \caption{(a) Dependence of the vector potential amplitude of a surface mode with mode function proportional to $\frac{q}{\sqrt2}e^{-qz}$ (cf.~Eq.~\ref{surface_mode}) on wavenumber $q$ plotted at height above the surface $z/a=2$. 
    (b) Effective transition coefficient between A and B sites $t^{eff}$ for different distances $z$ from the surface, for $A_0 = 0.5$. The modulation is caused by the time-averaged Peierls phase. (c and d) Eigenvalues of Hamiltonian with surface mode $A_0=0.5$ perpendicular to the chain, mapped by their projections onto bare Hamiltonian's eigenvectors at $z$ values depicted in (b), with `1' denoting 100\% share of lower band and `0' denoting 100\% share of the upper band. The grey lines indicate the bands in the absence of the electromagnetic field. (c) Band mixing as a consequence of the effective hopping coefficient $t^{eff}$ turning negative. (d) Weak field regime for comparison.}
    \label{fig:perp}
\end{figure*}

To this end, we model the saw-tooth chain to be in the vicinity of a substrate hosting surface phonon polaritons.
For the mode functions we assume the lower half space to be filled with an insulator, giving rise to a frequency- and space-dependent permittivity described by a Lorentz oscillator model for a single phonon resonance,
\begin{equation}
    \varepsilon(\omega, \vec r) = \begin{cases}
        &1 \hspace{23mm} ; \,z> 0\\
        &\varepsilon(\omega) = \frac{\omega_{\rm LO}^2 - \omega^2}{\omega_{\rm TO}^2 - \omega^2} \hspace{2mm} ; \, z < 0
    \end{cases}.
\label{eq:Permittivity}
\end{equation}
Here $\omega_{\rm LO}$ and $\omega_{\rm TO}$ are longitudinal and transversal phonon frequencies, respectively.
For frequencies $\omega$ within the restrahlenband $\omega_{\rm TO} < \omega < \omega_{\rm LO}$ the permittivity is negative, $\varepsilon(\omega) < 0$, such that no light can travel in the material within this frequency window, giving rise to perfect reflectivity.
Within the restrahlenband, p-polarized surface modes of the electromagnetic field appear that are evanescent modes on both the material and vacuum sides, and are thus exponentially localized near the interface.
Their mode-functions are given by 
\begin{equation}
    \begin{aligned}
        f_{>}(\vec r) &= \frac{1}{N} \left(\frac{q_x}{|\vec q|}, \frac{q_y}{|\vec q|}, i \sqrt{|\varepsilon(\omega)|} \right)^{\rm T} e^{- \frac{|\vec q| z}{\sqrt{|\varepsilon(\omega)|}}} e^{- i \vec q \cdot \vec r}\\
        f_{<}(\vec r) &= \frac{1}{N} \left(\frac{q_x}{|\vec q|}, \frac{q_y}{|\vec q|}, i \frac{1}{\sqrt{|\varepsilon(\omega)|}} \right)^{\rm T} e^{\sqrt{|\varepsilon(\omega)|} |\vec q| z} e^{ i \vec q \cdot \vec r},
    \end{aligned}
    \label{eq:CavityQED_WFSurface}
\end{equation}
where $\vec r$ and $\vec q$ are in-plane position and wave vector, respectively.
Here $f_{>}$ and $f_{<}$ denote the mode functions above and below the interface, respectively.
In order to obtain orthonormal modes in the case of a frequency- and space-dependent permittivity $\varepsilon(\omega, \vec r)$ the normalization constant $N$ has to be fixed by
\begin{equation}
    \int_V \mathrm{d}\vec r \, \frac{\varepsilon(\vec r, \omega_q)}{2} \left(1 + \frac{1}{\varepsilon(\vec r, \omega_q)} \frac{\partial (\omega_q \varepsilon(\vec r, \omega_q))}{\partial \omega_q}\right) |\vec f_q(\vec r)|^2 = 1,
\label{eq:CavityQED_NormalizationDispersing}
\end{equation}
where $V$ is the quantization volume. For the surface modes this yields
\begin{equation}
    N^2 = \frac{(1 + |\varepsilon(\omega)|) \left(|\varepsilon(\omega)| + \frac{ \left[ 1 + \omega_{\rm TO}^2 \frac{\omega_{\rm LO}^2 - \omega_{\rm TO}^2}{\left(\omega^2 - \omega_{\rm TO}^2\right)^2} \right]}{|\varepsilon(\omega)|} \right)}{2 |\vec q| \sqrt{|\varepsilon(\omega)|}}.
\end{equation}
Effects from the surface modes can be well understood in the large $|\vec q|$ limit, where their dispersion is essentially flat, and one has $\varepsilon(\omega) \approx -1$.
For this case, subsuming the constants and considering only the in-plane part of the electromagnetic field above the interface due to the surface modes and remembering that the overall field is real, we thus write
\begin{equation}
    {\bf A}_{\text{surf}}({\bf r})=A_0\sum_{\bf q}\frac{\sqrt{|\vec q|} \vec e_q}{\sqrt{2}}e^{-qz} \cos(\bf{qr} - \omega t),
    \label{surface_mode}
\end{equation}
with $\vec e_q = \frac{1}{|\vec q|} (q_x, q_y)^{\rm T}$ the in-plane polarization along the wave-vector.

\section{Results}

\subsection{Perpendicular field}

\begin{figure*}[htp!]
    \centering
    \includegraphics{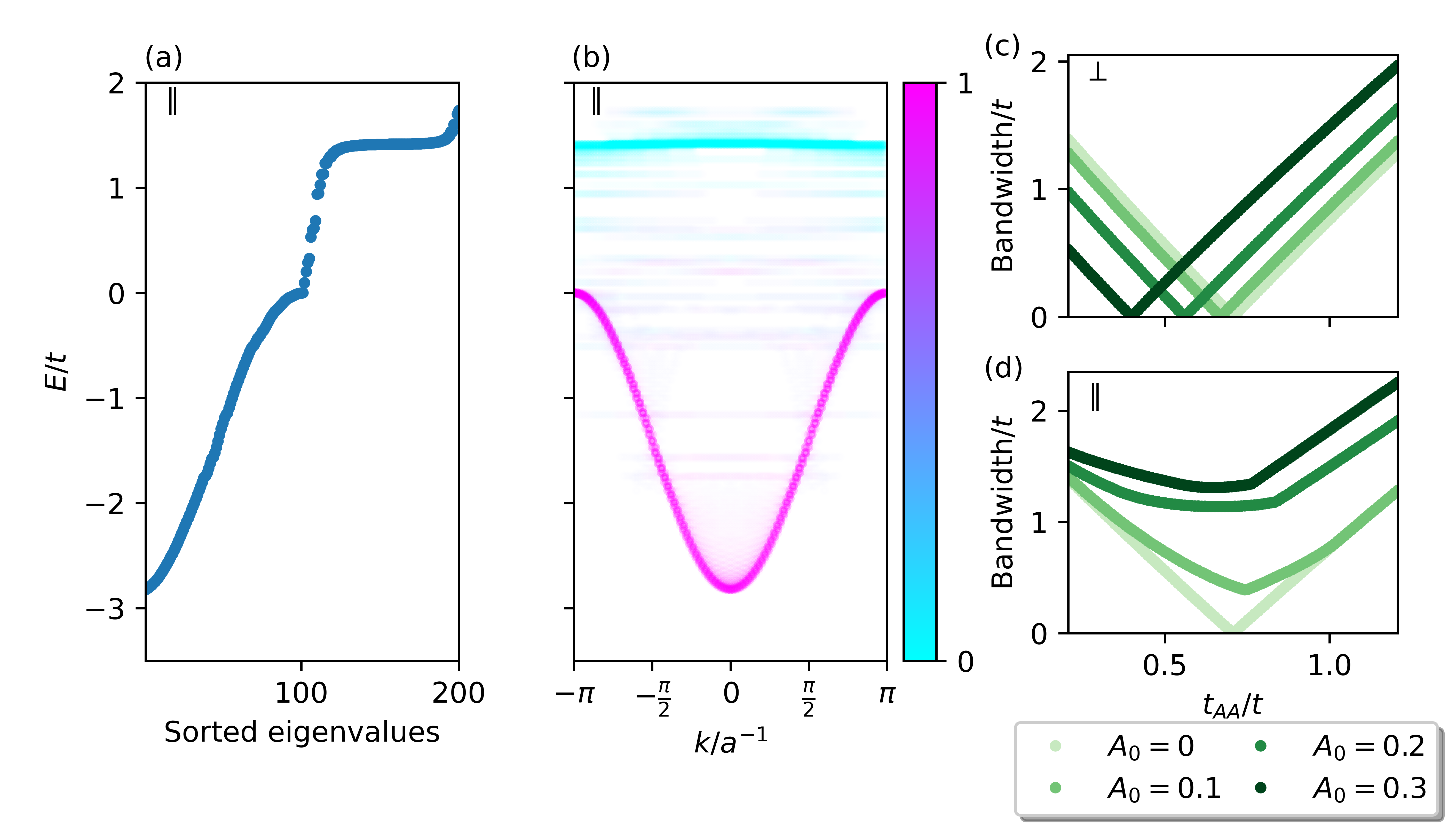}
    \caption{(a) Eigenvalues of Hamiltonian with surface mode parallel to the chain, $A_0 = 2$, $z=1$, sorted in ascending order. (b) The same eigenvalues mapped by their corresponding eigenvectors' projection onto bare Hamiltonian's eigenvectors, with `1' denoting 100\% share of lower band and `0' denoting 100\% share of the upper band. (c) Bandwidth of the upper band as a function of $t_{AA}$ hopping, for different perpendicular field strengths. (d) Same as in (c) but for field applied parallel to the chain. Bandwidth of the upper band is defined here as the difference between the maximum and median eigenvalues of the entire two-band spectrum. All results are for $N=100$ unit cells (200 sites including sublattice).}
    \label{fig:par}
\end{figure*}

We study the effect of applying a field perpendicular to the axis of the chain, i.e., along the $y$ direction in Fig.~\ref{fig:flat}(a). This polarization involves no momentum transfer along the chain, hence leaves the momentum label $k$ intact. 

We first consider a spatially uniform ($q=0$) field, i.e., setting ${\bf A}_q = (A_0 \delta({\bf q}),0,0)$ in Eq.~\ref{eq:tdepvector}. We consider two different situations: In Fig.~\ref{fig:flat}(b), we start from an exactly flat upper band and show that the homogeneous perpendicular field can unflatten this flat band. In Fig.~\ref{fig:flat}(b), we instead start with a non-flat upper band, which is then flattened by field. The basic mechanism explaining these observed effects is that the field couples differently to the $t_{AA}$ and $t$ hopping terms of the chain, thus renormalizing them in a different fashion. Therefore the field allows to tune the ratio between the hopping matrix elements and thus the curvature specifically of the nearly flat band. 

However, as already mentioned above, reaching the required high field intensities in the far field is quite unrealistic. We therefore now turn to the discussion of the more realistic scenario of driven surface modes, Eq.~\ref{surface_mode}. In reality such surface modes exist with both in-plane polarizations $x$ and $y$. We envision that a linearly polarized laser can drive these modes selectively, coupling to the near field via, e.g., a scanning tunneling microscope tip. 

We first investigate again the situation of a drive polarized perpendicular to the chain, Fig.~\ref{fig:perp}. The vector potential's momentum profile is shown in Fig.~\ref{fig:perp}(a). 
Since the light intensity does not vary in the direction of the chain, all of the $t$ coefficients are modified in the same way, and the system can be regarded as a sawtooth chain with different transition rates between sites $A$ and $B$, which we call $t^{eff}$. This effect depends sensitively on the strength of the field, and can even cause the sign of $t$ to change. The strength of the surface mode field, on the other hand, depends strongly on the distance between the surface and the sawtooth chain. The dependence of the relative change of the hopping coefficient, $t^{eff}/t$, on the height above the surface $z$ is shown in Fig.~\ref{fig:perp} (b).
$t^{eff}$ depends on the field amplitude through the Bessel function, and the field amplitude decays exponentially with distance from the surface. This leads to oscillations near $z=0$ and a plateau at long distances, as the field strength decays to zero and $t^{eff}/t \rightarrow 1$, recovering the unrenormalized limit. Importantly, we note that reaching the field strengths required to reverse the sign of Floquet-renormalized hoppings in free space is virtually impossible; by contrast, in the near-field setting such strong renormalizations are not out of the question.

We now turn to the discussion of the implications of such strongly renormalized effective hoppings on the resulting light-matter engineered bands. In regions where the effective hopping coefficient is negative the original bands strongly mix in the central part of the first Brillouin zone, as shown in Fig. \ref{fig:perp} (c). This gives rise to unflattening of the originally flat band and a decrease of the overall band width of the system, making both bands have similar bandwidths.
On the other hand, in regions with positive $t^{eff}$ the bands only bend slightly without mixing, again most visibly near the centre of the Brillouin zone (Fig. \ref{fig:perp} (d)).

\subsection{Parallel field}

We now investigate the effects of applying light parallel to the chain, which leads to momentum transfer between the surface modes and the electrons in the chain. In this case the notion of a well-defined band structure becomes blurry, as electronic momentum ceases to be a good quantum number. Therefore the model is no longer solvable in closed form, and we provide numerical diagonalization results for a finite system size. The results of applying a surface mode (\ref{surface_mode}) parallel to the chain are presented in Fig. \ref{fig:par} (a), (b). Fig.~\ref{fig:par} (a) presents the resulting eigenstates as a sorted list of eigenvalues. Importantly, emergence of states within the energy gap between the bands of the bare Hamiltonian can be observed. This can be directly compared to Fig.~\ref{fig:par} (b), which presents the eigenvalues projected onto the bare Hamiltonian's eigenstates, with shading corresponding to the share of the lower band state. This figure reveals that there is significant mixing of states with different momentum coming from the same band (as expected from the Hamiltonian (\ref{eq:curvature}), and that the in-gap states result from mixing of the two bands. 

Finally we turn to the discussion of the degree of flexibility over bandwidth control that the two distinct driving schemes, perpendicular versus parallel to the chain, can offer. We define effective bandwidth for the upper, flat band as the energy difference between the maximum of energy and the median of the sorted list, since in the bare Hamiltonian the upper half of the energy spectrum belongs mainly to the flat band. Fig.~\ref{fig:par}(c) and (d) show the comparison of the effects of applying a surface mode perpendicular versus parallel to the chain on the bandwidth of the flat band, for different starting values of $t_{AA}$, and for varying field strengths. As can be seen in Fig.~\ref{fig:par} (c), the hopping parameter can be tuned to give a perfectly flat band for any starting value of $t_{AA}$ when the field is applied perpendicular to the chain. Fig.~\ref{fig:par} (d), on the other hand, shows that applying the field parallel to the chain mainly broadens the band, due to the emergence of in-gap states. 

\section{Discussion}
In this work we have demonstrated that flexible Floquet control over a two-band model sawtooth chain is possible by coupling it to the surface polariton modes of a suitable substrate. We have shown that this allows one to modify the band dispersions and specifically the effective band width of a nearly flat band, an effect that is made possible by quantum-geometric light-matter coupling \cite{topp_light-matter_2021}. Moreover we have discussed the impact of the choice of light polarization on the resulting band structures. Importantly, the quasi-one-dimensional nature of the chain allows one to tune the light polarization perpendicular to the chain, which enables flexible control by flattening and unflattening the nearly flat band while retaining sharply defined band structures. On the other hand, tuning the light polarization parallel to the chain generically leads to band broadening and lifts the conservation of electronic momentum. This opens the path to controlling finite-momentum scattering channels depending on the momentum structure of the surface polaritons.

The interplay between electron-polariton scattering and coupling of electrons to other modes, such as phonons, will open new playgrounds to control quantum materials with polaritonic fields. A particularly interesting route specifically in nearly flat-band materials is the interplay of light-matter coupling and correlation effects, such as the ones observed in moir\'e or kagome materials, among others. As an example, the influence of surface light-matter coupling on different types of charge orders in kagome metals \cite{neupert_charge_2022} is an interesting route to explore new light-matter control paradigms, akin to the recent demonstration of cavity control over the charge density wave transition in 1$T$-TaS$_2$ \cite{jarc_cavity-mediated_2023}. It will be particularly interesting to study the impact of driven surface polaritons on flat-band superconductors \cite{peotta_superfluidity_2015}, or the potential to induce superconductivity with polaritons instead of laser light \cite{mitrano_possible_2016}. 

We finally note that beyond changing effective bandwidths and the resulting densities of states, the effective hopping renormalization also leads to mixing of bands, which as a consequence affects their resulting quantum geometry and ultimately also their topology if band inversion happens. This opens additional tuning paths for Floquet-like topological control over quantum materials \cite{rudner_band_2020}. 

\section{Acknowledgments}
MAS was funded by the European Union (ERC, CAVMAT, project no. 101124492).
\bibliography{flattening}
\bibliographystyle{apsrev4-2}

\appendix
\section{Derivation of the light-matter coupling}
We start from a vector potential ${\bf A}({\bf r},t)$ with components
\begin{equation}
    A^\mu ({\bf r},t) = \sum_{\bf q}A_{\bf q}^\mu(t) e^{i{\bf qr}}.
\end{equation}
Introducting a general tight-binding Hamiltonian with Peierls phase factors,
\begin{equation}
\begin{aligned}
    H &= \sum_{ab}\sum_{ij} -t_{ab}(i,j)c_{i,a}^\dagger c_{j,b}e^{i\int_{R_j,b}^{R_{i,a}}dr'_\mu A^\mu({\bf r'})},
\end{aligned}
\end{equation}
the integral is approximated by taking the value at the midpoint times the distance between endpoints.
\begin{equation}
\begin{aligned}
    H&\approx 
    \sum_{ab}\sum_{ij} -t_{ab}(i,j)c_{i,a}^\dagger c_{j,b} e^{iA^\mu(\frac{{\bf R_i+R_j}}{2})(R_{i,a\mu}-R_{j,b\mu})}\\ 
    &\approx
    \sum_{ab}\sum_{ij} -t_{ab}(i,j)c_{i,a}^\dagger c_{j,b}\\&\times (1 + iA^\mu(\frac{{\bf R_i+R_j}}{2})(R_{i,a\mu}-R_{j,b\mu}) \\&- \frac12 [A^\mu(\frac{{\bf R_i+R_j}}{2})(R_{i,a\mu}-R_{j,b\mu})]^2).
\end{aligned}
\end{equation}
This approximation is valid for situations relevant here, in which the spatial variations (wavelength) of the light field is orders of magnitude larger than the atomic length (size of unit cell, or lattice constant).

The discrete Fourier transform is introduced in a standard way:
\begin{equation}
\begin{aligned}
    H&= \sum_{ab}\sum_{ij}\sum_{{\bf kk'}} -t_{ab}(i,j)c_{{\bf k},a}^\dagger c_{{\bf k'},b} (1 \\&+ i\sum_{\bf q}e^{i{\bf q}\frac{\bf R_i + R_j}{2}}(R_{i,a\mu}-R_{j,b\mu})A_{\bf q}^\mu \\ &- \frac12\sum_{\bf qq'}[e^{i{\bf q}\frac{\bf R_i + R_j}{2}}e^{i{\bf q'}\frac{\bf R_i + R_j}{2}}(R_{i,a\mu}-R_{j,b\mu})\\&\times A_{\bf q}^\mu(R_{i,a\nu}-R_{j,b\nu})A_{\bf q'}^\nu\\
    &+ e^{i{\bf q}\frac{\bf R_i + R_j}{2}}e^{-i{\bf q'}\frac{\bf R_i + R_j}{2}}(R_{i,a\mu}-R_{j,b\mu})\\&\times A_{\bf q}^\mu(R_{i,a\nu}-R_{j,b\nu})A_{\bf q'}^{\nu *}] + H.c.\\ 
    \end{aligned}
\end{equation}

It is assumed that the hopping amplitudes depend only on the difference vector between the sites, and not their positions, and so the $j$ index is replaced with ${\bf \Delta R}$:
\begin{equation}
\begin{aligned}
H&= \sum_{ab}\sum_{i}\sum_{\bf kk'}\sum_{\bf \Delta R} -t_{ab}({\bf \Delta R})c_{{\bf k},a}^\dagger c_{{\bf k'},b} \{ e^{i{\bf (k-k')R_i}}e^{i{\bf k'\Delta R}}\\ 
&- i\sum_{\bf q}e^{i{\bf (k-k'+q)R_i}}e^{i{\bf (k' + \frac{q}{2})}}\Delta R_\mu A_{\bf q}^\mu \\ &- \frac12\sum_{\bf qq'}[\Delta R_\mu A_{\bf q}^\mu \Delta R_\nu A_{\bf q'}^\nu e^{i{\bf(k-k'+q+ q')R_i}}e^{i{\bf (k'+\frac{q}{2} + \frac{q'}{2})\Delta R}}\\ 
&+ \Delta R_\mu A_{\bf q}^\mu \Delta R_\nu A_{\bf q'}^{\nu *} e^{i{\bf (k-k'+q- q')}}e^{i{\bf (k'+\frac{q}{2} - \frac{ q'}{2})\Delta R}}] \} \\
&=\sum_{ab}\sum_{\bf k} \{c_{{\bf k},a}^\dagger c_{{\bf k},b} \sum_{\bf \Delta R} -t_{ab}(\bf \Delta R)e^{i{\bf k\Delta R}} \\&+ i\sum_{\bf q} c_{{\bf k},a}^\dagger c_{{\bf k+q},b} \sum_{\bf \Delta R}-t_{ab}({\bf \Delta R})A_{\bf q}^\mu \Delta R_\mu e^{i{\bf (k + \frac{q}{2})\Delta R}} \\&- \frac12\sum_{\bf qq'}[c_{{\bf k},a}^\dagger c_{{\bf k+q+q'},b}\\&\times \sum_{\bf \Delta R}-t_{ab}({\bf \Delta R})\Delta R_\mu A_{\bf q}^\mu \Delta R_\nu A_{\bf q'}^\nu e^{i{\bf (k + \frac{q}{2} + \frac{ q'}{2})\Delta R}} \\&+ c_{{\bf k},a}^\dagger c_{{\bf k+q-q'},b}\\&\times \sum_{\bf \Delta R}-t_{ab}({\bf \Delta R})\Delta R_\mu A_{\bf q}^\mu \Delta R_\nu A_{\bf q'}^{\nu *} e^{i{\bf (k + \frac{q}{2} - \frac{q'}{2})\Delta R}}]\}+H.c.\\
\end{aligned}
\end{equation}
Writing this more concisely by introducing the dispersion relation of the bands, and noticing that higher order terms are derivatives of the unperturbed dispersion relations at different values of ${\bf k}$, we arrive at (\ref{eq:curvature})
\begin{equation}
\begin{aligned}
H&=\sum_{ab}\sum_{\bf k}\{c_{{\bf k},a}^\dagger c_{{\bf k},b}t_{ab}({\bf k}) \\&+ \sum_{\bf q}c_{{\bf k},a}^\dagger c_{{\bf k+q},b}A_{\bf q}^\mu \partial_{k_\mu}t({\bf k+\frac{q}{2}})
\\&+ \frac12\sum_{\bf qq'}[c_{{\bf k},a}^\dagger c_{{\bf k+q+q'},b}\\&\times A_{\bf q}^\mu A_{\bf q'}^{\nu}\partial_{k_\mu}\partial_{k_\nu}t_{ab}({\bf k+\frac{q}{2} + \frac{{\bf q'}}{2}})
\\&+ c_{{\bf k},a}^\dagger c_{{\bf k+q-q'},b}\\&\times A_{\bf q}^\mu A_{\bf q'}^{\nu *}\partial_{k_\mu}\partial_{k_\nu}t_{ab}({\bf k+\frac{q}{2} - \frac{ q'}{2}})] \} +H.c.   
\end{aligned}
\end{equation}
\end{document}